\documentclass{style/svproc}

\usepackage{url}

\usepackage{graphicx}
\usepackage{amsmath}
\usepackage{physics}
\usepackage{multirow}
\usepackage{booktabs}
\usepackage[caption=false]{subfig}

\newcommand{\eg}{\textit{e}.\textit{g}., }

\begin{document}
\mainmatter

% ----------------------------------------------------------

\title{Technologies for AI-Driven Fashion Social Networking Service with E-Commerce}
\titlerunning{AI-Driven Fashion SNS with E-Commerce}

% ----------------------------------------------------------

\author {
    Jinseok Seol \inst{1} \and
    Seongjae Kim \inst{1} \and
    Sungchan Park \inst{2} \and
    Holim Lim \inst{1} \and
    Hyunsoo Na \inst{1} \and
    Eunyoung Park \inst{2} \and
    Dohee Jung \inst{3} \and
    Soyoung Park \inst{3} \and
    Kangwoo Lee \inst{3} \and
    Sang-goo Lee \inst{1}
}
\authorrunning{Jinseok Seol et al.}
\tocauthor {
    Jinseok Seol,
    Seongjae Kim,
    Sungchan Park,
    Holim Lim,
    Hyunsoo Na,
    Eunyoung Park,
    Dohee Jung,
    Soyoung Park,
    Kangwoo Lee,
    Sang-goo Lee
}

\institute{
    Seoul National University, Department of Computer Science and Engineering \\
    \email{\{jamie, sjkim, ihl7029, monchana, sglee\}@europa.snu.ac.kr}
\and
    IntelliSys Co., Ltd. \\
    \email{\{scpark, eypark\}@intellisys.co.kr}
\and
    LOTTE Homeshopping Inc. \\
    \email{\{conan94, soyong.park19, kangwoo.lee\}@lotte.net}
}

% ----------------------------------------------------------

\maketitle

% ----------------------------------------------------------

% ==============================================================================

% ----------------------------------------------------------

\begin{abstract}
    The rapid growth of the online fashion market brought demands for innovative fashion services and commerce platforms.
    With the recent success of deep learning, many applications employ AI technologies such as visual search and recommender systems to provide novel and beneficial services.
    In this paper, we describe applied technologies for AI-driven fashion social networking service that incorporate fashion e-commerce.
    In the application, people can share and browse their outfit-of-the-day (OOTD) photos, while AI analyzes them and suggests similar style OOTDs and related products.
    To this end, we trained deep learning based AI models for fashion and integrated them to build a fashion visual search system and a recommender system for OOTD.
    With aforementioned technologies, the AI-driven fashion SNS platform, \textit{iTOO}, has been successfully launched.
\end{abstract}

% ----------------------------------------------------------

\section{Introduction}

    With the development of the internet and computer technologies, the size of the e-commerce market has been growing steeply.
    Moreover, the social distancing environment of the COVID-19 pandemic has brought the growth and demands for innovative e-commerce platforms \cite{silvestri2020future}.
    % To meet the demands and achieve benefits, leading commerce companies such as Amazon \cite{linden2003amazon, shrestha2019deep}, eBay \cite{yang2017visual, wang2021personalized}, Alibaba \cite{zhang2018visual, wang2018billion} are introducing distinctive and novel services, including visual search and product recommendations.
    To meet the demands and achieve benefits, leading commerce companies such as Amazon \cite{shrestha2019deep}, eBay \cite{yang2017visual}, Alibaba \cite{zhang2018visual} are introducing distinctive and novel services, including visual search and product recommendations \cite{mohanty2020recommender,jain2020exploiting}.
    Meanwhile, in the fashion industry, a variety of innovative applications have emerged.
    For example, ViSENZE \cite{chokshi2020transforming} provides fashion image processing solutions including image search and attribute prediction, and Zalando research team has been working for fashion product recommendation \cite{freno2017practical}.
    There are more interesting applications such as Intelistyle \cite{zou2021fashion} which provides a chatbot-based AI stylist, and a wardrobe-based AI stylist Fitzme \cite{park2019deep}.

% --- begin figure ---
\begin{figure}[ht]
    \centering
    \includegraphics[width=0.65\textwidth]{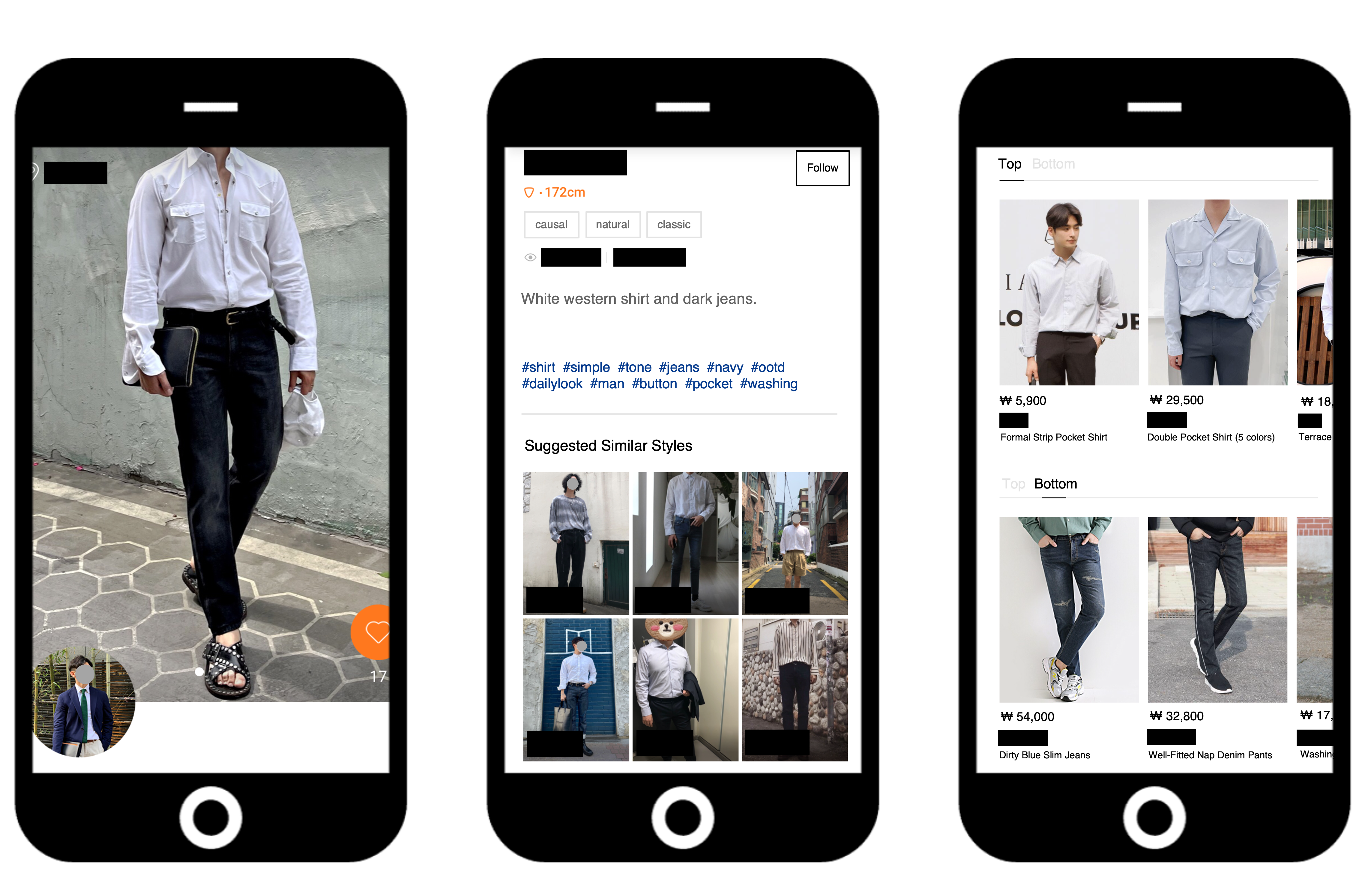}
    \caption{Actual usage screen of \textit{iTOO}, a Fashion SNS from LOTTE. Whenever users upload their outfit-of-the-day (OOTD) photos, AI analyzes them and suggests similar style OOTDs and related products.}
    \label{fig:detail}
\end{figure}
% --- end figure ---

    Fashion-focused social networking is another large portion of consumer activities \cite{nelson2019fashion}.
    We claim that people look for the outfit-of-the-day (OOTD) of other people to acquire insights and trends in fashion.
    To achieve this, consumers browse online applications such as Lookbook, Polyvore, Pinterest, etc.
    Moreover, users commonly share their OOTD photos through general social networking services (SNS) (\eg Facebook, Instagram, and Twitter).
    In this environment, connecting fashion SNS with e-commerce is undoubtedly beneficial, as in the case of Instagram and Styleshare.
    Therefore, it is reasonable to come up with a new kind of service that incorporates e-commerce with fashion SNS by applying AI technologies.

    Normally, to connect user-uploaded OOTD photos with retail products, the uploader must directly attach a link to the product, which is a cumbersome task, leading to the automation challenge.
    We applied the fashion visual search system to overcome the challenge, thus the users can easily share their OOTD in the form of a common fashion SNS.
    It provides an opportunity for other users to purchase retail products similar to the OOTD, without burdening the uploader.
    To this end, we implemented AI components including a fashion object detector, a fashion category classifier, and a fashion attribute tagger.
    Moreover, we deployed a personalized recommender system that can handle OOTD data to engage more users to the application.
    These AI technologies are combined and enabled launching the \textit{iTOO}.
    
    % The following paper is structured as follows: In Chapter 2, studies related to AI technologies of the application are explained.
    % Chapter 3 gives a brief overview of the deployed application \textit{iTOO} from LOTTE Homeshopping. Chapter 4 explains the AI components.
    % The main system, fashion visual search, and the recommender system are explained in Chapters 5 and 6, respectively.
    % Chapter 7 describes the system deployment configurations.
    % Finally, chapter 8 summarizes the paper with future works.

% ----------------------------------------------------------

\section{Related Work}

    \subsection{Deep Learning for Fashion}
    
        The most important media type in the fashion domain is photographic images.
        However, it contains major challenges that make it difficult to process images in the fashion domain \cite{cheng2021fashion}.
        Typically, self-occlusions may occur in the target of interest, and when a person is wearing fashion garments, the image variance can be amplified due to the viewpoint, posture, lighting, and scale of the subject \cite{zhan2017cross}.
        In addition, several fashion items may appear in a single image simultaneously.
        Therefore, localization procedure is essential.
        To this end, region-of-interest (RoI) detection, landmark detection \cite{liu2016fashion}, parsing with a fashion component \cite{liang2015deep}, or pose estimation \cite{toshev2014deeppose} is often employed.
        Moreover, classifying the categories of fashion products \cite{cho2019leveraging}, predicting colors and detailed attributes including latent fashion style, is also a major component to recognize fashion item \cite{li2016human,lee2017style2vec}.
        % Fashion information can be extracted and recognized from photos only through numerous processes.
    
    \subsection{Visual Search System}
    
        Image retrieval, or a visual search system, has been successfully applied in various areas such as face recognition \cite{schroff2015facenet} and product search \cite{yang2017visual}.
        Recently, many studies implemented a visual search model by comprehending images through CNN and learning through deep metric learning \cite{chen2021deep}.
        Academic datasets for fashion visual search are publicly available \cite{liu2016deepfashion,ge2019deepfashion2}, but to apply in the real-world application, datasets should cover a broader range of categories.
        Therefore, collecting and refining the dataset is also an essential task \cite{corbiere2017leveraging}.
        % On the other hand, to train the visual search model, the same-class label indicating that different images of the same instance are required.
        Techniques using proxies by employing latent embeddings for each instance are the current state-of-the-art models for the visual search \cite{kim2020proxy}, to the best of our knowledge.
        However, when the number of items becomes millions, training causes another challenge and requires complex techniques \cite{zhao2019large}.
    
    \subsection{Recommender System}
    
        The recommender system is a core technology in contents platforms and e-commerce as Amazon \cite{linden2003amazon}, YouTube \cite{covington2016deep}, and Netflix \cite{bennett2007netflix} have shown.
        Starting with Collaborative Filtering \cite{resnick1994grouplens,sarwar2001item}, advanced methods including implicit-based approaches \cite{hu2008collaborative}, content-based filtering \cite{basu1998recommendation}, and more recently, deep learning-based recommendation algorithms \cite{he2017neural} have been studied.
        In the fashion domain, models using visual information have been mainly studied \cite{he2016vbpr,yin2019enhancing}.
        Furthermore, there are many cases that multiple images form a single outfit, thus outfit recommendation techniques that consider multiple images simultaneously are being studied \cite{lin2020outfitnet,lu2021personalized}.

% ----------------------------------------------------------

% --- begin figure ---
\begin{figure}[ht]
    \centering
    \includegraphics[width=0.65\textwidth]{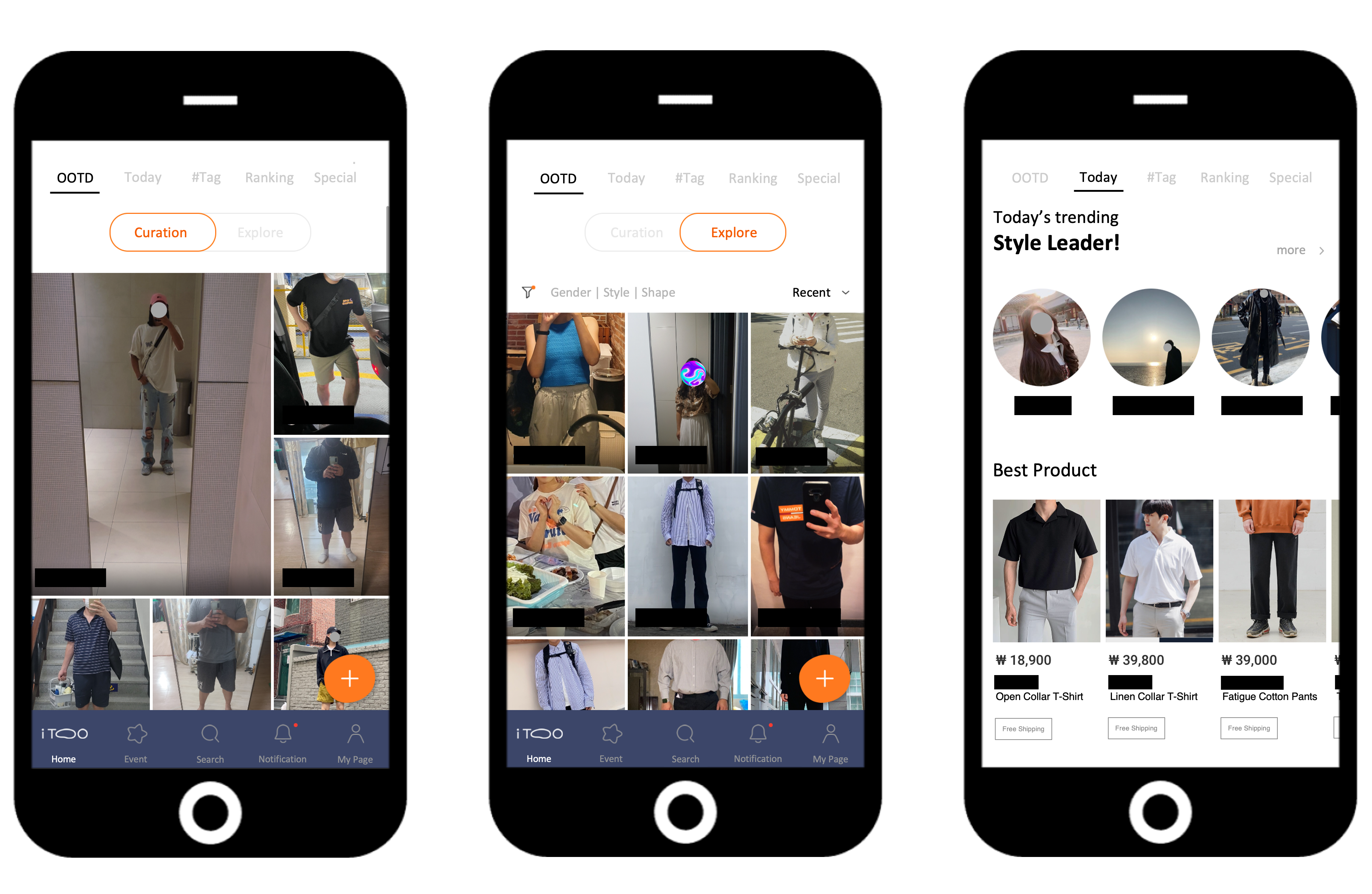}
    \caption{The home screen of our AI-driven fashion SNS, \textit{iTOO}. Users can browse and interact with the OOTDs of other users just like in ordinary SNS. Most of the contents are curated by the recommender system.}
    \label{fig:home}
\end{figure}
% --- end figure ---

\section{Service Overview}

    In this section, we introduce \textit{iTOO} from LOTTE.
    \textit{iTOO} is a service that integrates Fashion SNS and commerce, where you can share and browse your OOTD photos, look for related products and styles, get recommendations, and even purchase fashion products in place.
    
    \subsection{Share and Browse OOTD}
    
        One of the main purposes of the application is to share a picture of your OOTD.
        A user can add a brief description and hashtags when uploading the picture.
        Immediately, the fashion items that comprise the OOTD are detected and analyzed automatically by AI.
        Therefore, users can easily share extensive information by simply taking a picture and leaving a short description.
        Moreover, users can browse through OOTDs posted by other users through curation or exploration.
        As shown in Figure \ref{fig:home}, the home screen recommends OOTDs that fit the preference, style, and body shape of the user.
    
    \subsection{Look into OOTD}
    
        By examining the OOTD detail view, a user can check out purchasable retail products that are similar to the comprising items of the OOTD as illustrated in Figure \ref{fig:detail}.
        This function benefits users who want to buy fashion products through OOTD curation in place, and retail shops can merchandise their product through viral marketing.
        In addition, other OOTDs with a similar style are recommended. Such curations help users to drill down OOTD pools that fit the preference of the user.
    
    \subsection{Get More Recommendations}
    
        Besides aforementioned OOTD recommendations, the ``style leaders'' who often post trending and decent OOTDs are also recommended to a user as who-to-follow.
        To get more accurate curations, a user can provide detailed information of the fashion persona such as demographic information, body shape, and preference style tags.
        All information including OOTD interactions, profile, following user list is gathered to recommender system and provides a personalized recommendation.

% ----------------------------------------------------------

\section{AI Components}
    
    This section introduces the AI technologies that enable fashion SNS focused on OOTD images.
    We used the best performing deep learning models in our knowledge, and they were fine-tuned to be suited for the fashion domain and the application.
    The core models of the application, visual search and OOTD recommender system, are described in Sections 5 and 6 respectively.
    Note that the part of the AI component is also used to construct datasets for training the other AI components.
    
    \subsection{Fashion Object Detector}
        \label{sec:detector}
    
        A fashion image may present a single product, but in many cases, it comes with a person wearing several fashion garments.
        Therefore, localizing or detecting where the fashion items are in the image is a process that must be preceded.
        We considered pose estimation and human parsing methods, however, we adopted a model that predicts region of interest (RoI) for practicality because human information does not always come in.
        As a training dataset, we mixed and reorganized Street2Shop \cite{hadi2015buy}, ModaNet \cite{zheng2018modanet}, and DeepFashion \cite{liu2016deepfashion,ge2019deepfashion2} datasets.
        In detail, we mapped category information to 6 super-categories (top, bottom, outer, dress, shoes, bag) in order to combine datasets from different sources.
        % For the detector model, the family of Mask R-CNN model \cite{he2017mask} and Yolo model \cite{redmon2016you} were take into account.
        Due to the throughput performance issue, YoloV4 \cite{bochkovskiy2020yolov4} were selected as our detector model.
        Note that in the application situation, we could assume that each OOTD image has at least one fashion item.
        Moreover, top/bottom items and dresses are mutually exclusive, so considering these properties, we added a post-processing module and achieved performance gain in terms of recall.
        Furthermore, we also included the fashion category classifier module from Section \ref{sec:classifier} to increase the precision.
        The predicted RoI is cropped and inferred by the category classifier, and filtered out if the super-category is different from the detector model.
        By combining the YoloV4 with the post-processing module, we could build a fast and accurate fashion object detector.
    
    \subsection{Fashion Category Classifier}
        \label{sec:classifier}
    
        Classifying the category of a fashion item is another basic element of fashion item recognition.
        We constructed an integrated dataset using ModaNet, DeepFashion, iMaterialist \cite{guo2019imaterialist} and crawled data from YOOX and Polyvore.
        To increase category coverage, we crawled the data from the top popular 30 online fashion malls, summing up to 1.7M images in total.
        Since the crawled data does not have RoI labels, we used the fashion object detector model from Section \ref{sec:detector} and filtered out RoIs with a super-category label parsed from metadata provided by the malls.
        It is necessary to reorganize the category hierarchy to integrate multiple datasets, thus we designed a category hierarchy consisting of 6 super-categories and 32 sub-categories: 6 from outer, 6 from top, 6 from bottom, 2 from dress, 7 from shoes, and 5 from bag.
        As a classification model backbone, EfficientNet \cite{tan2019efficientnet} was employed under consideration of inference speed and memory usage.
        We also leverage the training techniques such as cosine annealing and label smoothing, etc.

% % --- begin table ---
% \begin{table}[]
%     \centering
%     \caption{Fashion attribute tags obtained from Fashion Attribute Tagger. The upper 7 attribute groups are multi-label, and the bottom 11 are categorical. Some attribute groups have a category constraint, where they are filtered out through post-processing according to the inferred category of the input image.}
%     \label{tab:attribute}
%     \begin{tabular}{l|r|l}
%         \toprule
%         Attribute Group & Tags & Constraint \\
%         \midrule
%         Style        & 10   &              \\
%         Adjective    & 16   &              \\
%         Season       & 4    &              \\
%         Occasion     & 4    &              \\
%         Detail       & 25   &              \\
%         Fabric       & 16   &              \\
%         Pattern      & 10   &              \\
%         \midrule
%         Color        & 12   &              \\
%         Sleeve Type  & 4    & outer, top   \\
%         Neckline     & 11   & top          \\
%         Top Fit      & 3    & top          \\
%         Pants Length & 2    & bottom/Pants \\
%         Pants Fit    & 5    & bottom/Pants \\
%         Jeans Color  & 4    & bottom/Jeans \\
%         Skirt Shape  & 6    & bottom/Skirt \\
%         Skirt Length & 3    & bottom/Skirt \\
%         Dress Length & 4    & dress        \\
%         Heel         & 4    & shoes        \\
%         \bottomrule
%     \end{tabular}
% \end{table}
% % --- end table ---

    \subsection{Fashion Attribute Tagger}
        \label{sec:tagger}
    
        We build a fashion attribute tagger model to find detailed attributes of fashion items such as color, style, length.
        Specifically, we defined 18 attribute groups where 11 are categorical, and 7 are multi-label.
        Since some attribute groups are only limited according to the sub-category, outputs are filtered out through the post-processing module.
        Similar to other AI component models, we merged and reorganized multiple datasets from different sources: DeepFashion, iMaterialist, Fashion550k \cite{inoue2017multi}, and MVC \cite{liu2016mvc}.
        Adopted CNN backbone and training methods are the same as fashion category classifier.

% ----------------------------------------------------------

\section{Fashion Visual Search}

    To train the visual search model, the same-class labels denoting different images of the same item are necessary.
    In the fashion domain, the image variance especially the gap between the product image provided by the shopping malls and the image uploaded by consumers is large.
    Therefore, it is important to construct a model and datasets that can cover the cross-domain image retrieval task.
    To this end, we collected multiple datasets and pre-processed them to build an in-house dataset suitable for the fashion visual search model in the application.
    Note that since it is common to learn through negative sampling rather than learning the distribution of all items, dataset quality is sensitive to false-positives rather than false-negatives.
    
    \subsection{Dataset Construction}
    
        \subsubsection{Collecting Data}
        
            Academic datasets (\eg DeepFashion) are often not complete and cannot be directly applied to real-world applications due to the limitation of category coverage.
            To fill this gap, we selected and crawled the top popular 30 online fashion malls and collected a total of 0.3M items with 1.3M images, including consumer photo review data.
            We conducted a small experiment to confirm that adding more data affects search performance in terms of category coverage.
    
        \subsubsection{Preprocessing}
        
            Similar to the case of the fashion category classifier from Section \ref{sec:classifier}, crawled data cannot be used for training without preprocessing.
            We used the fashion object detector model from Section \ref{sec:detector} and acquired RoI crops for the localization, and filtered out the crops that do not match the super-category information parsed from target malls.

            Meanwhile, when crawling the data from online malls, the abundant image data is often located in the ``descriptive image'', which consists of multiple photos of a fashion item, description texts, and even irrelevant images like advertisements.
            To gather meaningful data, we first separated the descriptive image with the connected components algorithm, then removed duplicate images using perceptual hashing \cite{zauner2010implementation}.
            The detector model and post-process procedures are applied then after.
            Additionally, to reduce false-positive errors, we use the fashion category classifier and select images with the sub-category of the majority.
    
    \subsection{Color Separation}
    
        An easy-to-miss aspect when building a dataset for a fashion visual search model is to separate fashion items that have multiple color variants.
        Many online fashion malls, including DeepFashion dataset, treat item images that differ only in color as the same item.
        However, this scheme can lead CNN to neglect the color information of the input image, and a ``shortcut'' by color information cannot be used.
        In our settings, it is beneficial to use this shortcut because the precision of the search result is more important than the recall, and by conducting a benchmark experiment, we confirmed that separating the color variants into different items helps to improve precision.
        In the case of the DeepFashion dataset, the color information labels are provided with fine granularity, so we re-adjusted the same-class label using the color tag of our fashion attribute tagger from Section \ref{sec:tagger}.
        Again, when it comes to precision, only the false-positives of the dataset matter so the inaccuracy of the attribute tagging model does not affect critically.
    
    \subsection{Model}
    
        Most of the recent state-of-the-art methods on image retrieval tasks are based on metric learning.
        When the model is trained, we can obtain a representation vector from the input image by feeding it into the model, and similar items can be retrieved through cosine similarity.
        We considered basic metric learning \cite{schroff2015facenet}, AP learning \cite{cakir2019deep} and proxy-based methods \cite{kim2020proxy}.
        However, methods that require item embeddings are often difficult to deal with numerous or variable item pool.
        Moreover, we use an under/over-sampling scheme to balance the datasets from different sources, which means, the whole item pool is changed on every epoch.
        Therefore, for the flexibility of the training, we adopted simple $N$-pair contrastive learning \cite{sohn2016improved}.
        Although the basic metric learning cannot match the state-of-the-art performance, it still serves as a decent baseline with advantages from other aspects.

        In concrete, to train the $N$-pair loss, we sample one positive (same-item) image per input image and gather $N$ negative image samples from the training batch.
        The metric learning is performed using normalized-temperature cross-entropy (NT-Xent) loss \cite{chen2020simple}.
        The rest of the training detail including backbone CNN is similar to the category classifier from Section \ref{sec:classifier}.
        The dimension of the representation vector was set to 128 for memory efficiency, and although a larger dimension was under consideration, the performance improvement compared to memory usage was not significant.
        The under/over-sampling of datasets are empirically adjusted considering the image types, characteristics, and category distribution of each dataset.

% --- begin table ---
\begin{table}[ht]
    \caption{Performance comparison on DeepFashion In-shop dataset, using top-$k$ accuracy. Our $N$-pair based model may not be state-of-the-art, but it can be easily scaled out to millions of items. The suggesting color separation training scheme (last row) shows that even with simple label modification, the top-1 accuracy can be improved by a large margin.}
    \label{tab:deepfashion}
    \centering
    \begin{tabular}{l|cccc}
    
        \toprule
        
            Model
            & $k$=1  & $k$=5  & $k$=10  & $k$=20  \\
            
        \midrule
        
            Liu et al. 2016b
            & 53.0  & -     & 73.0  & 76.0  \\
            
            Park et al. 2019
            & -     & 82.6  & -     & 90.9  \\
            
            Cakir et al. 2019
            & 90.9  & -     & 97.7  & 98.5  \\
            
            Kim et al. 2020
            & 92.6  & -     & 98.3  & 98.9  \\
        
        \midrule
        
            Baseline
            & 77.9  & 91.6  & 94.5  & 96.5  \\
            
            Color Separation
            & 83.4  & 92.4  & 94.0  & 95.6  \\
            
        \bottomrule
    \end{tabular}
\end{table}
% --- end table ---

    \subsection{Experimental Results}
    
        \subsubsection{Performance on Benchmark Dataset}
        
            To show the precision gain on the color separation scheme, we experimented on DeepFashion In-shop dataset, which has 52,712 images with 7,982 items.
            Note that this dataset provides RoI crop data, so we use the box coordinates with 20-pixel margin, then resized it into 256 $\times$ 256 images.
            Table \ref{tab:deepfashion} shows the results in top-$k$ accuracy, which checks whether the positive image is within the top-$k$ items retrieved.
            Although the $N$-pair baseline model cannot reach state-of-the-art performance, it is still a decent baseline compared to older and complex models.
            On the other hand, we can see the significant performance gain in top-1 when the color separation scheme is applied.

% --- begin figure ---
\begin{figure}[ht]
    \centering
    \includegraphics[width=0.9\textwidth]{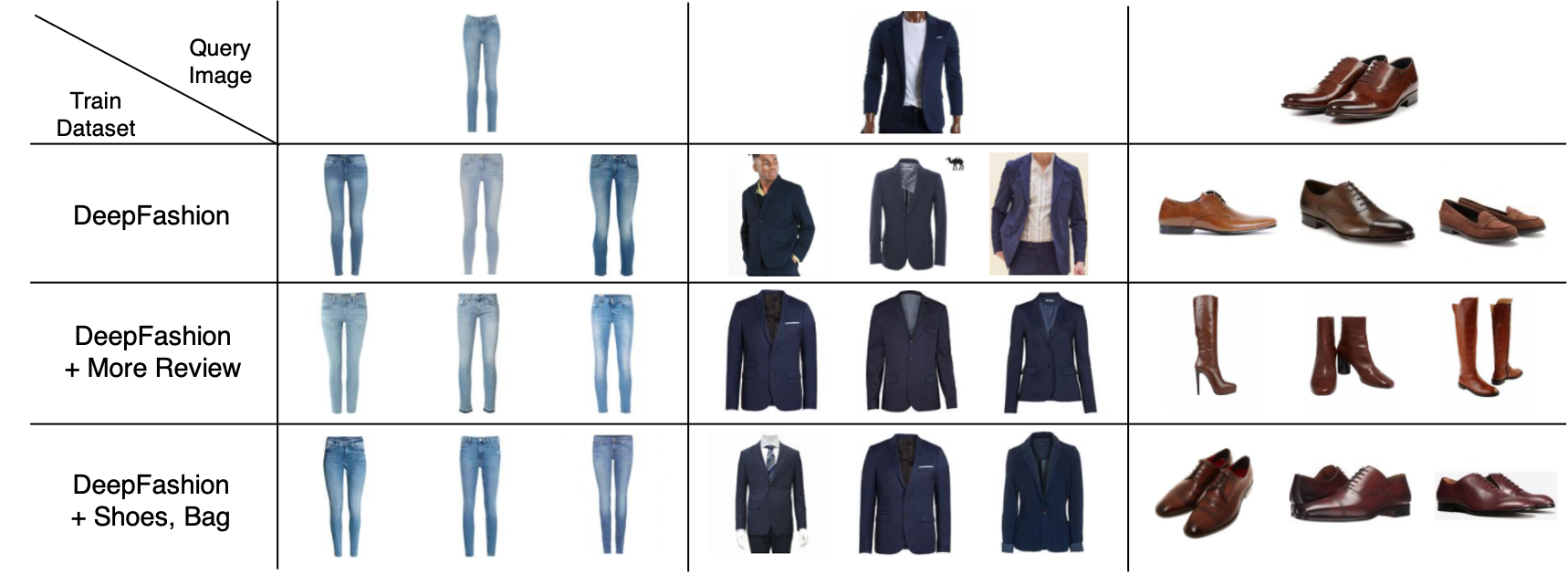}
    \caption{Examples of results from visual search models, trained in different dataset compositions. Results from the second query show that by adding more review data, robustness to cross-domain retrieval can be improved. The third query shows that the model cannot accurately deal with unseen categories. We can conclude that in the visual search model, dataset composition is critical as model architecture.}
    \label{fig:query}
\end{figure}
% --- end figure ---

% --- begin figure ---
\begin{figure}[ht]
    \centering
    \includegraphics[width=0.9\textwidth]{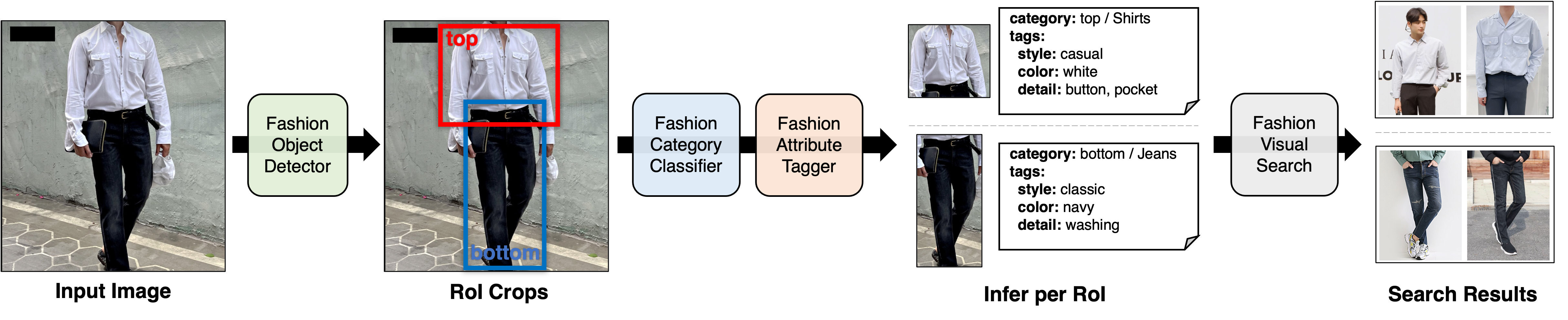}
    \caption{Inference pipeline for OOTD. After the fashion object detector model finds region of interest (RoI), the fashion category classifier and the fashion attribute tagger are applied to acquire more detailed information for each cropped RoI. After that, the visual search model extracts the representation vectors and stores them to the corresponding vector index.}
    \label{fig:pipeline}
\end{figure}
% --- end figure ---

        \subsubsection{Dataset Influence}
        
            To see the influence of the dataset constitutions, we trained a $N$-pair model with three different dataset settings: using only DeepFashion dataset, adding more consumer photo review data from crawled online fashion malls, and adding shoes and bags which DeepFashion does not have.
            The examples of visual search results are shown in Figure \ref{fig:query}.
            In the results from the first query image, all three settings produced similar results.
            In the second case, since the query image involves a partially human shape, the setting with an additional consumer review image shows more robustness in terms of cross-domain image retrieval.
            In the final case, where the query image represents shoes, it can be seen that settings without shoes and bags could not maintain the sub-category of the query image.
            As a result, we argue that constructing a well-tempered dataset is just as important as selecting the model.

    \subsection{Inference Pipeline}
    
        In the application, the visual search model has to be combined with other AI components.
        As shown in Figure \ref{fig:pipeline}, when a user uploads an OOTD image, the fashion object detector first finds RoIs
        Then, the fashion category classifier and the fashion attribute tagger are applied to each cropped RoI in parallel.
        Finally, the visual search model extracts representation vectors and store them into the vector index, corresponding to the super-category.

% ----------------------------------------------------------

\section{Recommender System}

    With the recommender system, users receive personalized OOTD recommendations, similar styled OOTDs for each OOTD, and style leaders to follow.
    % Through these recommendations, users can get content more suitable for them.
    
    \subsection{Personalized OOTD Curation}
    
        On the first screen of the service, users can get the recommendation of OOTDs that suits their preferences.
        The recommendation basically leverage CF-CBF, and the final recommendation list is generated by mixing up with the weekly best products and best products by demographic-based user segment.
        In the case of CF-CBF, both user-based and item-based CF are used.
    
        \subsubsection{Style Vector}
        
            A fashion item vector $\vb{v}_{i}$ of item $I_i$ consists of a concatenation of representations of the category classifier, the attribute tagger, and the visual search model: for an item image $x_{i}$,
            % \begin{equation}
                $\vb{v}_{i} = \text{concat}(f_{\text{C}}(x_{i}), f_{\text{A}}(x_{i}), f_{\text{S}}(x_{i})).$
            % \end{equation}
            For each fashion item, the item style vector $\tilde{\vb{v}}_{i}$ is obtained by subtracting the average of item vectors of the sub-category which the given item belongs: for $c(i)$ a sub-category index of an item $I_i$,
            % \begin{equation}
                $S_k = \{j \, | \, c(j) = k\},
            % \end{equation}
            % \begin{equation}
                \bar{\vb{v}}_{c(i)} = |S_{c(i)}|^{-1} {}\sum_{j \in S_{c(i)}} \vb{v}_{j},
            % \end{equation}
            % \begin{equation}
                \tilde{\vb{v}}_{i} = \vb{v}_{i} - \bar{\vb{v}}_{c(i)}.$
            % \end{equation}
            The OOTD style vector is defined as the average of the item style vector of the consisting items:
            % \begin{equation}
                $\vb{o}_{t} = |o^{*}_{t}|^{-1} \sum_{i \in o^{*}_{t}} \tilde{\vb{v}}_{i}$,
            % \end{equation}
            where $o^{*}_{t}$ is a set of indices of comprising items in the OOTD $o_t$.
    
        \subsubsection{Semantic OOTD Similarity}
        
            Given two OOTDs, we define semantic OOTD similarity as a weighted sum of the cosine similarity between the OOTD style vectors of each OOTD and the Jaccard similarity of the hashtags that are dependent on the two OOTDs: for given OOTDs $o_{t_{1}}$ and $o_{t_{2}}$,
            \begin{equation}
                \text{sim}_{\text{o}}(o_{t_{1}}, o_{t_{2}}) =
                    \lambda_{\text{o}}
                    \left(
                        \frac{
                            \vb{o}_{t_{1}}
                            \cdot
                            \vb{o}_{t_{2}}
                        }{
                            \left \| \vb{o}_{t_{1}} \right \|
                            \left \| \vb{o}_{t_{2}} \right \|
                        }
                    \right)
                    +
                    \left(1 - \lambda_{\text{o}}\right)
                    \frac{
                        |a_{\text{o}}(t_{1}) \cup a_{\text{o}}(t_{2})|
                    }{
                        |a_{\text{o}}(t_{1}) \cap a_{\text{o}}(t_{2})|
                    }
                ,
            \end{equation}
            where $a_{\text{o}}(t)$ denotes a set of hashtags of an OOTD $o_{t}$.
            Note that we use this similarity to make similar styled OOTD recommendations.
    
        \subsubsection{Semantic User Similarity}
        
            Similar to semantic OOTD similarity, we define a user style vector by aggregating style vectors of $H$ OOTDs that the user has recently viewed or liked.
            We use a weighted average to reflect the preferences of recent interaction more strongly: for user $u_{n}$,
            % \begin{equation}
                % $w_{m} = \left(\frac{H - m + 1}{H}\right)^\alpha$
            % \end{equation}
            \begin{equation}
                \vb{u}_{n} =
                    \left(\sum_{m}^H w_{m}\right)^{-1}
                    \sum_{m = 1}^{H}
                        w_{m}
                        \vb{o}_{t_{m}}
                ,
            \end{equation}
            where $w_{m} = \left(\frac{H - m + 1}{H}\right)^\alpha$, and $u^{*}_{n} = \{t_{1}, t_{2}, ..., t_{H}\}$ is a set of indices of the user's recent OOTD views or likes, and $0 < \alpha < 1$ is a recency decay hyper-parameter.
            Cosine similarity between two user style vectors and the Jaccard similarity between the preference tags in the user profiles are used to measure the semantic user similarity: for user $u_{n_{1}}$ and $u_{n_{2}}$,
            \begin{equation}
                \text{sim}_{\text{u}}(u_{n_{1}}, u_{n_{2}}) =
                    \lambda_{\text{u}}
                    \left(
                        \frac{
                            \vb{u}_{n_{1}}
                            \cdot
                            \vb{u}_{n_{2}}
                        }{
                            \left \| \vb{u}_{n_{1}} \right \|
                            \left \| \vb{u}_{n_{2}} \right \|
                        }
                    \right)
                    +
                    \left(1 - \lambda_{\text{u}}\right)
                    \frac{
                        |a_{\text{u}}(u_{n_{1}}) \cup a_{\text{u}}(u_{n_{2}})|
                    }{
                        |a_{\text{u}}(u_{n_{1}}) \cap a_{\text{u}}(u_{n_{2}})|
                    }
                ,
            \end{equation}
    
        \subsubsection{CF-CBF for OOTD}
        
            We use Collaborative Filtering (CF) as the basis for our recommendation algorithm.
            We first calculate the TF-IDF values from user-OOTD interactions.
            The value of TF-IDF is decayed to reflect the recency using time decay coefficient $\beta^d$, where $d$ is days passed since the interaction has occurred, and $0 < \beta < 1$ is a decay rate hyper-parameter.
            We use both item-based and user-based CF and combine it with other recommendation results.
            In the case of item-based, the recommended OOTD list is obtained through similarity of the OOTDs that the user has recently viewed.
            In the case of user-based, the recommendation list is constructed by joining users obtained through user similarity and the OOTD list that the user has recently viewed.
            In both cases, CF-CBF can be implemented by considering TF-IDF as a CF part and semantic similarity as a Content-Based Filtering (CBF) part.
            Let $\vb{r}_n$ be a TF-IDF vector of a user $u_{n}$, treating user as a document when calculating the TF-IDF values.
            Then, the final similarity between two users $u_{n_{1}}, u_{n_{2}}$ can be calculated as follows:
            \begin{equation}
                \text{sim}_{\text{CF-CBF}}(u_{n_{1}}, u_{n_{2}}) =
                    \lambda_{\text{CF}}
                    \left(
                        \frac{
                            \vb{r}_{n_{1}}
                            \cdot
                            \vb{r}_{n_{2}}
                        }{
                            \left \| \vb{r}_{n_{1}} \right \|
                            \left \| \vb{r}_{n_{2}} \right \|
                            + h
                        }
                    \right)
                    +
                    \left(1 - \lambda_{\text{CF}}\right)
                    \text{sim}_{\text{u}}(u_{n_{1}}, u_{n_{2}})
                ,
            \end{equation}
            where $h$ is a shrinkage term for the case with relatively small interactions \cite{bell2007improved}.
            Similar methodology is applied to item-based CF-CBF.
    
        \subsubsection{OOTD Curation}
        
            With user-based and item-based CF-CBF, we mix up the weekly best OOTD list with the best OOTD list by segment based on demographic information.
            Note that since a decay term is used, a result closer to global taste is provided rather than personalized content to those who have not used the application for a long time.
            This reflects the characteristics of the fashion domain where trends change over time and provides exploration oppurtunity and serendipity.

% --- begin figure ---
\begin{figure}[ht]
    \centering
    \includegraphics[width=0.8\textwidth]{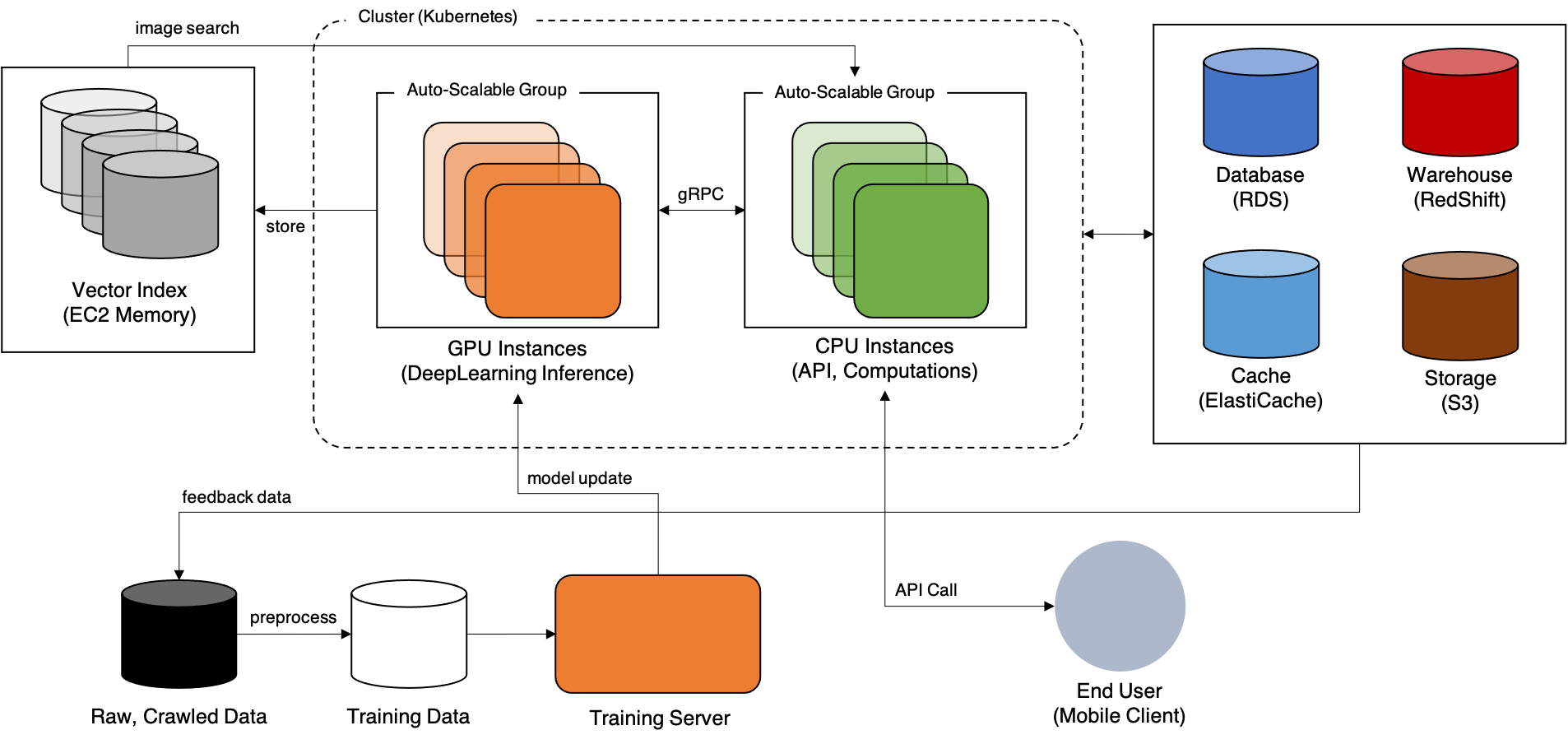}
    % \caption{Overall system architecture. To serve deep learning models in a real-world application, we adopted Kubernetes that can orchestrate complex infrastructure. Note that traffic-critical components are automatically scalable.}
    \caption{Overall system architecture. To serve deep learning models in a real-world application, we adopted Kubernetes that can orchestrate complex infrastructure.}
    \label{fig:system}
\end{figure}
% --- end figure ---

    \subsection{Style Leader Suggestion}
    
        A style leader means a person who can be subscribed, and a user can receive better OOTD curation when they follow the style leader.
        For style leader recommendation, both the latent method and the graph based method are used.
        In the case of the latent method, recommendation candidates are determined using the modified semantic user similarity.
        Here, we use cosine similarity between the user style vectors of the recent view/like OOTD history of the follower and the user style vectors of the recent \textit{upload} OOTD history of the followee candidates.
        On the other hand, when using the graph-based algorithm, recommendation candidates are obtained by performing a random walk twice in the following/follower relationship graph.
        Finally, we recommend a mixture of latent-based, graph-based, similar segment users using demographic information, and popular users.
        Segment and the weekly best serve as exploration and baseline at the same time.

% ----------------------------------------------------------

\section{System Deployment}

    \subsection{Overall Architecture}

        Serving deep learning models for real-world application requires high-cost and complex infrastructure.
        To minimize the burden, we adopt AWS Cloud Service, mainly orchestrated through Kubernetes.
        Deep learning models are loaded on auto-scalable GPU pods to adapt to the variable traffic.
        The overall architecture is illustrated in Figure \ref{fig:system}.

    \subsection{Serving Deep Learning Models}
    
        The development of deep learning models is usually done using frameworks such as PyTorch or TensorFlow with an experimental environment.
        To serve the trained model for the inference, we used NVIDIA Triton Inference Server since it can accommodate all types of neural network models exported in ONNX format, independent from the deep learning framework.
        For the communication between the service API and the deep learning models, we created an in-house gRPC client library.
        Each inference step is divided into CPU-heavy parts such as image preprocessing or data loading, and the core GPU consuming part so that each component can be scaled out in parallel.

    \subsection{Vector Indexing}
    
        Throughput of visual search system heavily relies on similar vector search algorithms \cite{liu2016deep}.
        We considered well-known approaches including Deep Hashing \cite{liu2016deep} and hierarchical search methods \cite{malkov2018efficient}.
        Empirically, vectors from visual search models form intrinsically clustered spaces, thus separating the hashing stage from the model does not degrade search accuracy compared to learning to hash methods.
        Therefore, HNSW \cite{malkov2018efficient} was adopted in consideration of implementation difficulty, search time complexity, and memory used.
        In our situation, hundreds of fashion items are added every day, so the vector index is rebuilt every dawn to include such items.
        In the case of HNSW, the memory consumption increases linearly for the items in the database.
        Therefore, whenever the index cannot be afforded by a single computing instance, we apply the sharding technique \cite{zhang2018visual,zhao2019large} and rearrange the search results through post-processing.
        Note that since the super-category of a fashion item is almost always accessible, vector indexes are built separately according to the super-categories.

    \subsection{Data Warehouse and DAG Management}
    
        When it comes to the recommender system, it is necessary to analyze logs and identify the user-item relationships from large-scale data.
        To this end, we implemented data processing modules and a basic CF model using AWS RedShift, the data warehouse instance.
        Moreover, both the visual search system and the recommender system are a pipeline of relatively small modules.
        In this structure, task parallelism can be applied to improve throughput.
        We adopted Argo as a Directed Acyclic Graph (DAG) task management tool to implement the task parallelism.
        Through Argo and Kubernetes configurations, we can automatically scale out the bottlenecks in the DAG.

% ----------------------------------------------------------

\section{Conclusion}

    In this paper, we describe technologies for AI-driven fashion SNS that incorporate fashion e-commerce.
    Users can share and browse their OOTD, while detailed fashion attribute analysis, similar products search, and getting recommendations are all automatically provided by AI.
    To this end, we built a fashion object detector, a fashion category classifier, a fashion attribute tagger, a fashion visual search system, and an OOTD recommender system.
    With all these techniques, the fashion SNS platform \textit{iTOO} from LOTTE has been launched.
    Future work is to tune the hyperparameters of the AI models and improve model architecture with user feedbacks.

% ----------------------------------------------------------

\section*{Acknowledgments}
    
This work was made in collaboration with Seoul National University, IntelliSys Co., Ltd., and LOTTE Homeshopping Inc.
Also, this work was partly supported by Institute of Information \& communications Technology Planning \& Evaluation (IITP) grant funded by the Korean government (MSIT) (No.2021-0-00302, AI Fashion Designer: Mega-Trend and Merchandizing Knowledge Aware AI Fashion Designer Solution).
Special thanks to Jeeseung Han.

% ----------------------------------------------------------

\bibliographystyle{bibtex/splncs03_unsrt}
\bibliography{reference}

% \begin{thebibliography}{6}

%     \bibitem {smit:wat}
%     Smith, T.F., Waterman, M.S.: Identification of common molecular subsequences.
%     J. Mol. Biol. 147, 195?197 (1981). \url{doi:10.1016/0022-2836(81)90087-5}
    
%     \bibitem {may:ehr:stein}
%     May, P., Ehrlich, H.-C., Steinke, T.: ZIB structure prediction pipeline:
%     composing a complex biological workflow through web services.
%     In: Nagel, W.E., Walter, W.V., Lehner, W. (eds.) Euro-Par 2006.
%     LNCS, vol. 4128, pp. 1148?1158. Springer, Heidelberg (2006).
%     \url{doi:10.1007/11823285_121}
    
%     \bibitem {fost:kes}
%     Foster, I., Kesselman, C.: The Grid: Blueprint for a New Computing Infrastructure.
%     Morgan Kaufmann, San Francisco (1999)
    
%     \bibitem {czaj:fitz}
%     Czajkowski, K., Fitzgerald, S., Foster, I., Kesselman, C.: Grid information services
%     for distributed resource sharing. In: 10th IEEE International Symposium
%     on High Performance Distributed Computing, pp. 181?184. IEEE Press, New York (2001).
%     \url{doi: 10.1109/HPDC.2001.945188}
    
%     \bibitem {fo:kes:nic:tue}
%     Foster, I., Kesselman, C., Nick, J., Tuecke, S.: The physiology of the grid: an open grid services architecture for distributed systems integration. Technical report, Global Grid
%     Forum (2002)
    
%     \bibitem {onlyurl}
%     National Center for Biotechnology Information. \url{http://www.ncbi.nlm.nih.gov}

% \end{thebibliography}

% ----------------------------------------------------------

% ==============================================================================

% ----------------------------------------------------------

\end{document}